# Spatio-temporal analysis of sprays by using Phase Doppler Anemometry data

Erika Rácz[a,*], Milan Malý[b], Jan Jedelský[b], Viktor Józsa[a]

[a] Department of Energy Engineering, Faculty of Mechanical Engineering, Budapest University of Technology and Economics, Műegyetem rkp. 3. H-1111 Budapest, Hungary

[b] Faculty of Mechanical Engineering, Brno University of Technology, Technicka 2896/2. 616 69 Brno, Czech Republic

**Abstract**

Spray characterization often relies on empirical formulas, statistical distributions, and derived quantities. Deterministic spray behavior originates from physics-governed mechanisms of atomization, e.g., nozzle geometry, boundary conditions, and hydrodynamic instabilities. Due to the stochastic nature of the atomization process, which originates from turbulence, chaotic perturbations, and droplet–droplet interactions, the temporal characteristics of dynamic behavior are seldom investigated. The combination of these processes leads to droplet clustering, which is a spatio-temporal behavior that is the focus of the current paper for an airblast atomizer. The measurement data by Phase Doppler Anemometry includes droplet size, velocity, and arrival time. Firstly, the theoretical and experimental interparticle time distributions are compared using a chi-squared hypothesis test, which concluded multimodality. Secondly, k-means clustering is applied to determine droplet clusters, whose number was determined by gap statistics. The above analysis was performed using an extensive database of various measurement positions, atomizing pressures, liquid preheating temperatures, and liquid types. It was found that cluster formation affects approximately 30% of the droplets in a single data set. In conclusion, the unsteadiness in the central region is caused by clustering, while it is caused by mixing and droplet entrainment in the spray periphery. The centroids and the number of cluster values depend on the atomizing pressure and the spray position, and are independent of the liquid temperature. The dynamical behavior of the clusters is compared by their droplet size and velocity distributions, showing no significant difference, suggesting that unsteady spray modeling is necessary if temporal characteristics are critical.

**Nomenclature**

Latin letters

| Notation | Description | Unit (if relevant) |
|---|---|---|
| $D$ | droplet size | μm |
| $j$ | bin width | ms |
| $An$ | error of the theoretical function | - |
| $F$ | droplet flux | 1/s |
| LHO | light heating oil | |
| max | maximum | |
| mean | mean value | |
| min | minimum | |
| $N$ | droplet number | - |
| $p$ | pressure | bar |
| $R$ | ratio of droplets in a cluster | % |
| RO | rapeseed oil | |
| std | standard deviation | |
| $t$ | arrival time of a droplet | ms |
| $T$ | total measurement time | μm |
| $T_p$ | preheating temperature | °C |
| $v$ | velocity | m/s |
| $V$ | Cramér's V | - |
| $x$ | radial distance on X axis | mm |
| X | axis with $x$ values | |
| $y$ | radial distance on Y axis | mm |
| Y | axis with $y$ values | |
| $z$ | axial distance on Z axis | mm |

Subscripts

| Notation | Description |
|---|---|
| $a$ | axial component |
| act | actual value |
| crit | critical value |
| g | property of gas phase |
| $k$ | number of the clusters |
| l | property of liquid |
| $r$ | radial component |
| $t$ | tangential component |
| th | theoretical function |
| exp | experimental data |

Greek letters

| Notation | Description | Unit (if relevant) |
| --- | --- | --- |
| $\alpha$ | significance level | - |
| $\Delta$ | difference | |
| $\lambda$ | intensity function | 1/ms |
| $\chi^2$ | chi-squared value | - |

# 1. Introduction

Several approaches are used to describe sprays in the literature, e.g., empirical formulas, statistical distributions, and derived quantities [1] due to the stochastic nature of the atomization process. However, deterministic effects, such as the inlet conditions, largely influence the spray properties, which makes certain characteristic quantities predictable. The combination of stochastic and deterministic effects results in coherent structures, which can be characterized by the correlated motions of the droplets. Hence, a time series analysis is required to completely characterize a spray. Therefore, the temporal behavior of airblast atomizer sprays is assessed in the present paper, which is rarely considered in spray models [1].

Although droplet clustering can occur even in steady sprays due to the random nature of droplet flux, in this discussion, the term clustering refers specifically to cases where the interparticle arrival times deviate in a statistically significant way from those expected under steady Poisson statistics. The unsteadiness of sprays arises from a complex interaction between hydrodynamic instabilities, injector-related effects, and external environmental factors. On the fluid-dynamic side, phenomena such as Rayleigh–Plateau [2] and Kelvin–Helmholtz instabilities, turbulence within the liquid jet or sheet [3], droplet-droplet interactions, including coalescence [4], and secondary breakup contribute to time-dependent variations in spray structure [1]. These processes are further modulated by gas–liquid coupling [5], where vortical structures [6] in the surrounding flow field induce additional fluctuations [7]. At the injector level, pressure oscillations [8,9], flow pulsations in the liquid supply system [10], cavitation within the nozzle [11], and geometric imperfections or deposits can contribute to unsteady processes. Environmental factors, such as turbulence and density gradients in the ambient gas [12,13], acoustic forcing [14,15] within confined chambers, and external vibrations or accelerations [16], may further amplify spray unsteadiness. Overall, spray unsteadiness is a multifactorial phenomenon, emerging from the combined action of internal instabilities, nozzle dynamics, and interactions with the surrounding flow field.

Spray unsteadiness is critical in, e.g., burners, where unsteadiness in either the atomization process or spray spreading [17] causes thermoacoustic problems, inhomogeneities in heat release, and increased pollutant emissions [18]. Consequently, spatio-temporal analysis is essential to address suboptimal operation in the design phase. In spray coating, unsteadiness results in inhomogeneous coverage and process control difficulties [19] due to uneven droplet arrival rate. In agricultural spraying and pesticide application, temporal fluctuations in droplet discharge result in heterogeneous pesticide deposition, with certain regions of the plant surface receiving excessive amounts while others are underdosed. Such non-uniformity decreases the overall efficacy of the treatment, promotes the development of pest resistance, and leads to excessive chemical consumption [20]. In pulmonary drug delivery, the aerosol dose must be both precise and reproducible. Spray pulsation introduces variability in drug deposition, thereby compromising dosing accuracy and reducing therapeutic effectiveness. Considering unsteadiness is essential for reliable measurements because temporal fluctuations affect both the reproducibility and interpretation of the results [21,22]. Ignoring unsteadiness can distort statistical characteristics, such as droplet size distribution or inter-arrival times, obscure important physical mechanisms, and even lead to an inappropriate choice of measurement techniques.

Edward and Marx [23–26] discussed the ideal spray theory. The ideal spray is defined to have four properties: the droplets of the spray are non-interacting point particles, each point particle carries a set of marks that represent the characteristics of the droplet, and the spray is treated as a random, Markov process. In this work, references to the homogeneity of the spray will be with respect to time. Additionally, the terms "inhomogeneous" or "unsteadiness" imply that the general ideal spray has non-uniform statistics in space and/or time. Assuming a steady spray, the interparticle arrival time distribution should follow Poisson statistics. Their multipoint statistical description enables the determination of the steadiness of the dispersed structures [26].

Different frameworks can be found in the literature on the investigation of spray steadiness. Phase Doppler Anemometry (PDA) enables the simultaneous measurement of droplet size, velocity, and arrival time; therefore, it is possible to conduct temporal analysis of the spray. The ideal spray theory has been tested successfully in several works investigating PDA data of, e.g., pressure, twin-fluid [27], effervescent [28,29], and a pressure atomizer sprays under swirling and hot conditions [30]. Jedelsky and Jicha [31] also applied this method to the PDA data of an effervescent atomizer. They concluded that the gas-to-liquid ratio influences the pressure fluctuations in the mixing chamber. Acharya et al. [32] investigated the process using spatio-temporal correlations and time-series modeling paradigms on PDA data of airblast and pressure-swirl atomizers. They employed a time-series framework, including autocorrelation and partial autocorrelation, to identify characteristic structures. Also, appropriate generalized autoregressive conditional heteroskedasticity models are developed. Heinlein et al. [33] employed neural network models fitted to the PDA data of a pressure atomizer, with a poor agreement between the experimental and calculated results in some spray regions due to unsteadiness and incomplete atomization.

The novelty of the present paper lies in the investigation of PDA data sets from an airblast atomizer to analyze the spatio-temporal behavior of the spray based on the interparticle arrival times of individual droplets. During the preceding visual investigations, irregularities were observed. Instead of the expected Poisson behavior in the interparticle arrival time distributions, multimodality was observed, suggesting the coexistence of several different physical phenomena and potentially distorting the quasi-steady nature of the spray with delayed droplet arrival. Omitting these irregularities may lead to significant bias. As a first step, the irregularities need to be decoupled to investigate their effect. For this purpose, k-means clustering was applied to the datasets, which estimates the number of possible droplet clusters. The elaborated calculation method utilizes a robust unsupervised learning algorithm that can be applied generally to different atomizers. The results contribute to the development of an advanced time-series model for spray data.

**2. Materials and methods**

The experimental setup is detailed in Section 2.1, which outlines the key parameters of the atomizer and the measurement conditions. Subsection 2.2 focuses on the processing of raw PDA data, with a particular emphasis on interparticle arrival times. All calculations were conducted in MATLAB R2024b software environment.

*2.1 PDA measurement*

The measurements and the atomizer summarized below were performed and detailed in earlier work [34]. This subsection emphasizes the key parameters to understand the processed data.

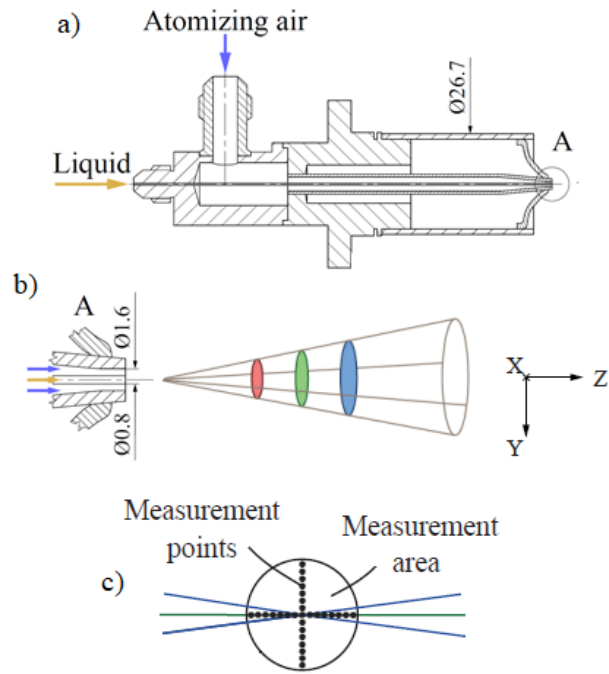

**Fig. 1.** Geometry of the airblast atomizer [34].

The cross-section of the evaluated airblast atomizer is shown in Figs. 1a and 1b. The liquid flows through a pipe with a 0.4 mm inner diameter while the atomizing air discharges from a concentric annulus (1.4 mm outer, 0.8 mm inner diameter). The high-velocity air blows over the surface of the low-velocity liquid jet, leading to a rapid disintegration of the liquid jet into ligaments, then into tiny droplets.

A two-component fiber-based PDA was used to measure droplet size and velocity. An Ar-ion laser with 514.5 and 488 nm wavelengths was used to measure the two velocity components, while the signal of the former beam pair was used to calculate the droplet diameter. The optical system was fixed while a computer-controlled 3D traverse system positioned the atomizer.

The PDA measurements were performed at 90 measurement points along the Z axis at three axial distances: z = 20, 40, and 60 mm, as shown in Figs. 1b and 1c. Along the X and Y axes at $z = 60$ mm, fifteen, at $z = 40$ mm, thirteen, and at $z = 20$ mm, seventeen equally spaced radial points were selected. At $z = 20$ mm, the step size was 1 mm between the points and 2 mm at $z = 40$ and 60 mm. The PDA system was limited to 40,000 individual particles or for a 15 s recording time, which rationally limited the acquisition in the peripheral regions.

Four different liquids were tested: diesel, light heating oil (LHO), rapeseed oil (RO), and water. Preheating temperatures, $T_p$, of 25, 40, 55, 70, and 100 °C were used, except for water, where the last point was 90 °C to avoid boiling. The following atomizing gauge pressures, $p_g$, were investigated: 0.3, 0.6, 0.9, 1.2, 1.8, and 2.4 bar. Table 1 contains all the measurement settings; each permutation was tested, resulting in 120 different sprays and 90 measurement points per spray. The range of the key non-dimensional numbers for the tested liquids and conditions is summarized in [35].

**Table 1.** The evaluated test conditions.

| Parameter | Value | | |
|---|---|---|---|
| Fuel | diesel. LHO. RO. water | | |
| $p_g$ [barg] | 0.3. 0.6. 0.9. 1.2. 1.8. 2.4 | | |
| $p_l$ [barg] | 0.01 | | |
| $T_p$ [°C] | 25, 40, 55, 70, 100 (water: 90) | | |
| $z$ [mm] | 20 | 40 | 60 |
| $x/y$ [mm] | -8–8 | -12–12 | -14–14 |
| Step [mm] | 1 | 2 | 2 |
| Measurement points along X/Y | 17 | 13 | 15 |

*2.2 Data processing*

Theoretically, the time distribution ($t$) of the droplet arrival times follows a Poisson behavior, while the interparticle arrival time distribution ($\Delta t$) follows an exponential distribution [24,26]. For the comparison of the hypothetical function and the experimental results, the following calculation was performed [27]. The interparticle time distribution of two consecutive droplets is:

$$\Delta t_i = t_i - t_{i-1}, \qquad (1)$$

at the time $i$, where $i \in [2, N]$ and $N$ is the total droplet number of a single measurement point. The experimental interparticle time function:

$$h_{\exp}(\tau_j) = \frac{H(\tau_j)}{N \cdot \Delta \tau_j}, \qquad (2)$$

where $\Delta \tau_j$ is the width of the $j^{\text{th}}$ interparticle time gap, $\tau_j$, and $H(\tau_j)$ is the number of $\Delta t_i$ that fall into that gap. The number of the interparticle time gaps ($j$) was calculated by the Rice rule:

$$j = 2 \cdot \sqrt[3]{N}. \qquad (3)$$

$\Delta \tau_j$ can be calculated by $j$ as follows:

$$\Delta \tau_j = \frac{T}{j}, \qquad (4)$$

where $T$ is the total measurement time. Based on [24], the theoretical interparticle time function, $h_{\text{th}}(\tau_j)$, normalized by the total number of the sample, and the event count, $H_{th}(\tau_j)$, is:

$$h_{\text{th}}(\tau_j) = \frac{(T - \tau_j) \cdot \lambda^2 \cdot e^{-\lambda \cdot \tau_j}}{\lambda \cdot T - 1 + e^{-\lambda \cdot T}}, \qquad (5)$$

$$H_{th}(\tau_j) = h_{\text{th}}(\tau_j) \cdot N \cdot \Delta\tau_j, \tag{6}$$

where $\lambda$ is the intensity function of the Poisson process, which is assumed to be constant over the measurement time and is calculated from the parameters of the experimental distribution:

$$\lambda = \frac{N}{T}. \tag{7}$$

Additionally, the theoretical function values are calculated for the mean of the limits of the bins of the experimental distribution.

## 3. Results and discussion

Firstly, the assessment of non-Poisson behavior and uncertainty is addressed. Also, finding the optimal number of clusters based on the results is emphasized in Subsection 3.1. In Subsection 3.2, the results of the $\chi^2$ test are summarized for the unsteadiness detection. Then, the results of the k-means clustering are shown in Subsection 3.3. After removing the additional clusters, unsteadiness detection was conducted again in Subsection 3.4. The comparison of the results of the liquids is summarized in Subsection 3.5. Then, the dynamical behavior of the droplet clusters is introduced in Subsection 3.6.

### *3.1 Assessment of non-Poisson behavior*

Figure 2 shows irregularities from the theoretical function in each presented case, while their source is unknown. The theoretical Poisson distribution has randomness: even if the spray were perfectly steady, finite sampling produces fluctuations. Firstly, it is necessary to determine whether the irregularities are due to random noise or evidence of real droplet clustering. In Fig. 2, the error bars show the expected deviation of the theoretical results due to the random nature of a theoretical spray. In each bin of the theoretical function, $H_{th}(\tau_j)$ is a discrete variable, which is the count of independent variables. Therefore, it follows a Poisson distribution with expected deviation equal to the square root of the count in each bin. Hence, the error is calculated as the deviation normalized with a linear transformation:

$$E_j = \frac{\sqrt{H_{exp}}}{N \cdot \Delta \tau_j} . \qquad (8)$$

If the difference exceeded what Poisson randomness could explain, the spray was genuinely unsteady, so the irregularity cannot be explained by randomness alone.

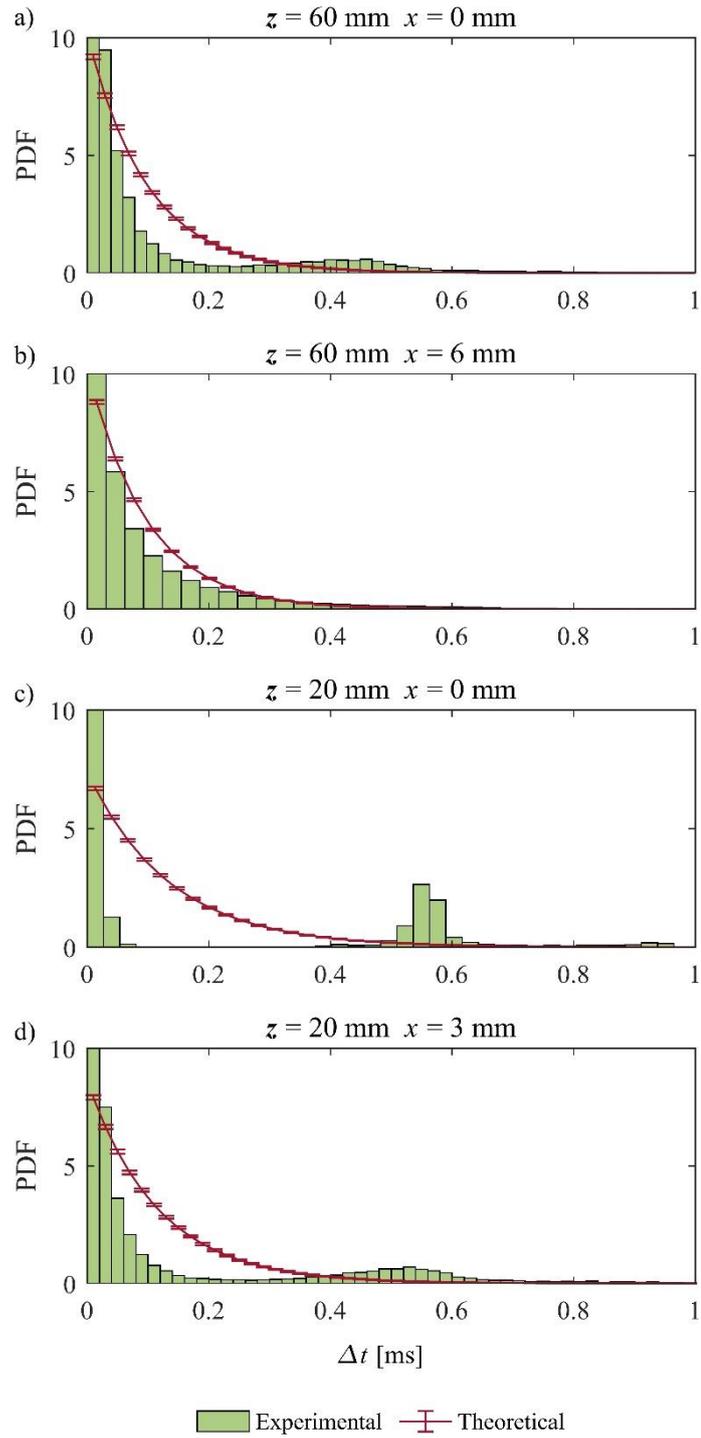

**Fig. 2.** Interparticle arrival time distributions for diesel at $T_p$ = 100 °C and $p_g$ = 1.8 bar: a) $z$ = 60 mm and $x$ = 0 mm, b) $z$ = 60 mm and $x$ = 6 mm, c) $z$ = 20 mm and $x$ = 0 mm, and d) $z$ = 20 mm and $x$ = 6 mm.

To prove steadiness or unsteadiness in a statistically rigorous method, the theoretical function and the experimental distribution are compared with the Chi-squared statistical hypothesis test; $\chi^2$ is calculated as:

$$\chi^2_{act} = \sum_{k=1}^{j} \frac{(h_{exp}(\tau_k) - h_{th}(\tau_k))^2}{h_{th}(\tau_k)}, \qquad (9)$$

and is assumed to follow a $\chi^2$ distribution with $(j-1)$ degrees of freedom.

The hypothesis that the experimental and theoretical values are similar can be accepted or rejected depending on the significance level, $\alpha$, which was selected at 0.05. The critical value of the statistics with $\alpha$ and $(j-1)$ degrees of freedom, $\chi^2_{crit}(1-\alpha, j-1)$ can be estimated for $\chi^2$ tables. If $\chi^2_{act} < \chi^2_{crit}$, the hypothesis can be accepted; therefore, steadiness can be expected at the local point of the spray, and the difference could be due to random fluctuations. Otherwise, it is rejected, and the spray is unlikely to be random, so it is unsteady. With a varying $j$ bin number, the value of $\chi^2_{act}$, and consequently, the test result may change significantly, as many datasets exhibit multimodal distributions, as shown in Subsection 3.1. Since the $\chi^2$ value depends on the sample size, Cramér's $V$ is used as a measure of effect size of the $\chi^2$ test of independence or goodness-of-fit:

$$V = \sqrt{\frac{\chi^2_{act}}{N \cdot (j-1)}}, \qquad (10)$$

which indicates the strength of the association (or difference) between two categorical distributions, regardless of sample size. In the case of relatively low $V$, around 0.15, the spray is not steady and statistically differs from the theoretical Poisson model; however, this effect is relatively small and practically insignificant. High $V$ indicates that the difference is significant in practical terms, for example, there is strong clustering in the spray, and the Poisson model results in a poor fit.

Taking this phenomenon into account, a further approach is needed to analyze the properties of multimodality that are not reflected in the results of the statistical tests. For determining the cluster number into which the droplets are divided, the k-means clustering algorithm was applied [36]. The k-means clustering works by choosing $k$ initial cluster centers, assigning each data point to the nearest center, then recalculating each center as the mean of its assigned points. This assign–update process repeats until the cluster assignments no longer change or a set number of iterations is reached, aiming to minimize the total Euclidean distance between points and their cluster centers.

During the visual investigations, the data showed clustering by $\Delta t$ into one to three clusters depending on the measurement parameters and the position in the spray. Therefore, the *evalclusters* [37] Matlab function was used to determine the optimal number of clusters based on gap statistics [38]. It compares the within-cluster dispersion of the clustering result against that of a reference dataset at the same range as the investigated data. The optimal $k$ cluster is chosen where the gap is maximized. Squared Euclidean distance was selected as a distance metric. For calculating the gap statistics, the reference data was generated as 100 uniformly distributed data points over the investigated $\Delta t$ range.

The droplets were partitioned into $k$ clusters, and each droplet belongs to the cluster with the nearest mean, called the centroids, $\Delta t_k$. The ratio of the droplet number in one cluster, $N_k$, compared to the total droplet number:

$$R_k = \frac{N_k}{N}, \quad (11)$$

where $k \in [1, 3]$ is the index of the droplet clusters. The first cluster ($k = 1$) has the lowest $\Delta t_k$, while $\Delta t_k$ increases with increasing $k$. Additionally, specific conditions were defined for the calculations, which stipulated that in multi-cluster data, the second and third clusters must contain at least 100 droplets. By using this condition, we can avoid the distorting effect of outliers, which may appear as individual clusters. If the condition is not fulfilled, the clustering method is repeated with a lower $k$ until the gap statistics find the optimal solution. During the evaluation of the results, it was observed that increasing $k$, decreasing $R_k$, and increasing $\Delta t_k$ occurred. However, in a few measurement datasets (less than 1% of all datasets), the clustering method failed to recognize clusters with these attributes. Therefore, in these cases, the calculations were performed until the required attributes were obtained.

An investigation and comparison of the $\Delta t_k$, $k$, $R_k$, and dynamic behavior of the droplet clusters were conducted. Droplet size, $D$, velocity, $v$, distributions and statistics of the clusters were compared, as well as their dependency on the measurement parameters (liquid, $p_g$, and $T_p$) and the position in the spray ($x$, $y$, $z$). The velocity of an individual droplet, $v$, is calculated by its velocity components along the X and Y axes, respectively:

$$v = \sqrt{v_a^2 + v_r^2} \, , \tag{12}$$

$$v = \sqrt{v_a^2 + v_t^2} \, , \tag{13}$$

where $v_a$ is the axial component, $v_r$ is the radial component along the X axis, and $v_t$ is the tangential component along the Y axis.

*3.2 Unsteadiness identification*

Figure 2 shows the experimental and theoretical interparticle arrival time distributions in different positions for diesel at $T_p = 100$ °C and $p_g = 1.8$ bar. The experimental distribution was obtained using Eq. (2), while the theoretical distribution was obtained using Eq. (6). The discrepancy between the theoretical distribution and the experimental distribution takes several forms. Figure 2b shows a unimodal experimental distribution of $\Delta t$, which slightly differs from the theoretical function. Figures 2a and 2d show bimodal $\Delta t$ distributions with a second mode around 0.4–0.5 ms. Figure 2c shows a multimodal distribution with a third mode around 0.9 ms, encompassing a range of 0.1 to 0.4 ms without any droplet. In the case of multimodality, clearly identifiable modes of interparticle arrival time are observed, suggesting clustering of the droplets. Additionally, the error bars ($E_j$) show a significant discrepancy between the theoretical and experimental results. Figure 2b shows a unimodal distribution, but the deviations also suggest unsteady behavior despite the lack of clustering. Therefore, it is necessary to examine whether the unsteadiness arises naturally from or as a result of clustering. The parameters of atomization and the position in the spray can modify the shape of the distributions and/or multimodality may appear; hence, the clustering tendency strongly depends on the conditions. These effects on the clustering are explained in Subsection 3.3.

Due to high $N$, even the smallest differences are significant, therefore the $\chi^2$ test rejected the null hypothesis in nearly all datasets for $\alpha = 0.05$. The $V$ value normalizes the $\chi^2_{act}$, and is a more appropriate measure of the test results. Although the droplet size and velocity exhibit more asymmetry along the X-axis than along the Y-axis [16, 17], the results are presented along the X-axis to show the actual flow pattern from the measurement results.

Figure 3a shows the $V$ results, and Fig. 3b shows the droplet flux, $F$, on the x–z plane for diesel at $T_p = 25$ °C and $p_g = 0.3$ bar. Close to the axis at $x = 0$ mm, especially near the nozzle exit, a large effect size is observed, caused by the appearance of clusters. Additionally, it is unclear whether this effect is enhanced by the chaotic, three-dimensional interaction between the liquid and gas at the nozzle outlet. $V$ rapidly decreases with increasing $|r|$, the minima are reached around $r = \pm 5$–$7$ mm, depending on $z$. In this zone, $F$ and $V$ do not correlate with each other. Then, $V$ increases towards the spray periphery due to the enhanced mixing and entrainment of droplets, which reduces the droplet flux. At every measurement parameter, the same trend was observed, but with varying magnitudes. The investigated twin-fluid atomizer in [27] showed similar trends for the $\chi^2$ value, but $V$ is not discussed in the work.

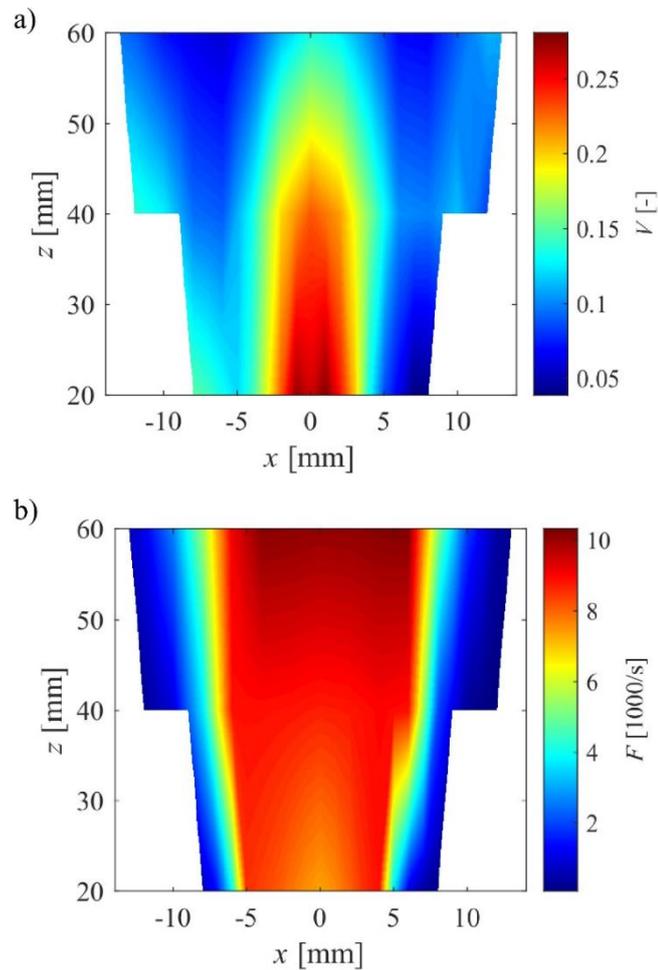

**Fig. 3.** Results of $V$ on the $x$–$z$ plane at $p_g = 0.3$ bar for diesel at $T_p = 25$ °C: a) effect size of the $\chi^2$ test, and b) the droplet flux, $F$, which is the ratio of the droplet number, $N$, and the total measurement time, $T$.

Figure 4 shows that $V$ depends on $p_g$ and $T_p$ along the X-axis for diesel at different $z$ values. As Figures 4a, 4c, and 4e show, with increasing $z$, $V$ decreases, while with increasing $p_g$, $V$ slightly increases. The $p_g$ = 1.8 and 2.4 bar are exceptions as supersonic flow [34] is created in the spray. Figures 4b, 4d, and 4f show that with increasing $T_p$, $V$ slightly increases around the $x$ = 0 mm central axis, but this effect disappears towards lower $z$.

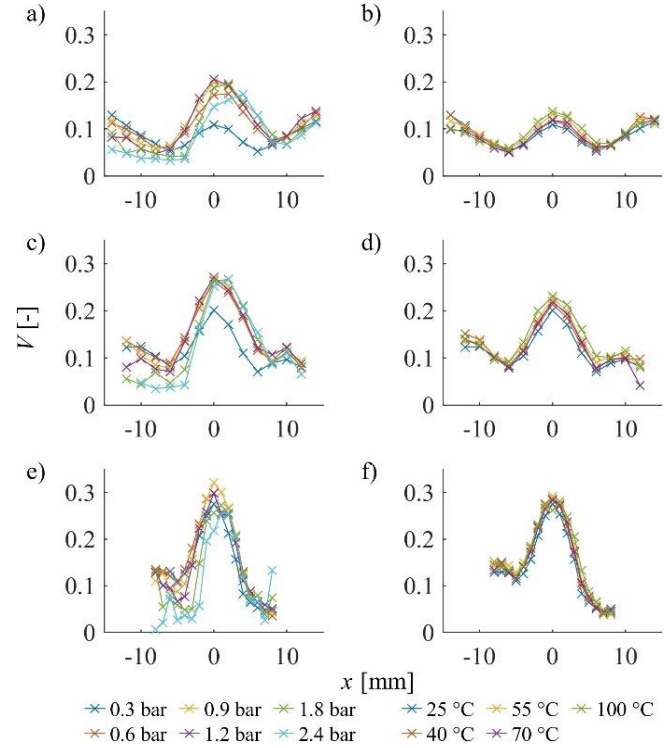

**Fig. 4.** $V$ depending on $T_p$ and $p_g$ along X axis for diesel: first column depending on $p_g$ at $T_p$ = 25 °C and second column depending on $T_p$ at $p_g$ = 0.3 bar, a–b) at $z$ = 60 mm, c–d) at $z$ = 40 mm, and e–f) at $z$ = 20 mm.

*3.3 Cluster Detection and Definition*

Figure 5 shows $\Delta t_k$ and $R_k$ of the clusts along the X-axis for diesel at $T_p$ = 25 °C and $p_g$ = 0.3 bar. Figures 5a, 5c, and 5e show the results for $\Delta t_k$ at different $z$. The presence of multiple clusters can be observed near the spray center, while the $k$ value decreases with both axial and radial distances. At the spray periphery, one cluster appears, where $\Delta t_1$ increases due to the low data rate and moderate droplet entrainment resulting from the relatively slower mixing process. $\Delta t_2$ and $\Delta t_3$ slightly decrease with increasing radial distances.

Figures 5b, 5d, and 5f show the corresponding $R_k$ of the clusters. $R_1$ decreases where the second and/or third clusters appear. As $z$ decreases, this effect is amplified. In conclusion, towards the spray core, where the atomization process is enhanced, the clustering effect is more likely to occur. Additionally, at higher $z$, in the fully developed region of the spray where the atomization process is complete, the second and third clusters tend to disappear or appear in the smaller $|r|$ region due to the rapid decrease in $F$ due to the expansion of the spray and turbulent mixing [39]. These characteristics also appear for all liquids, $p_g$, and $T_p$, although with varying intensities, which is discussed in Subsection 3.5.

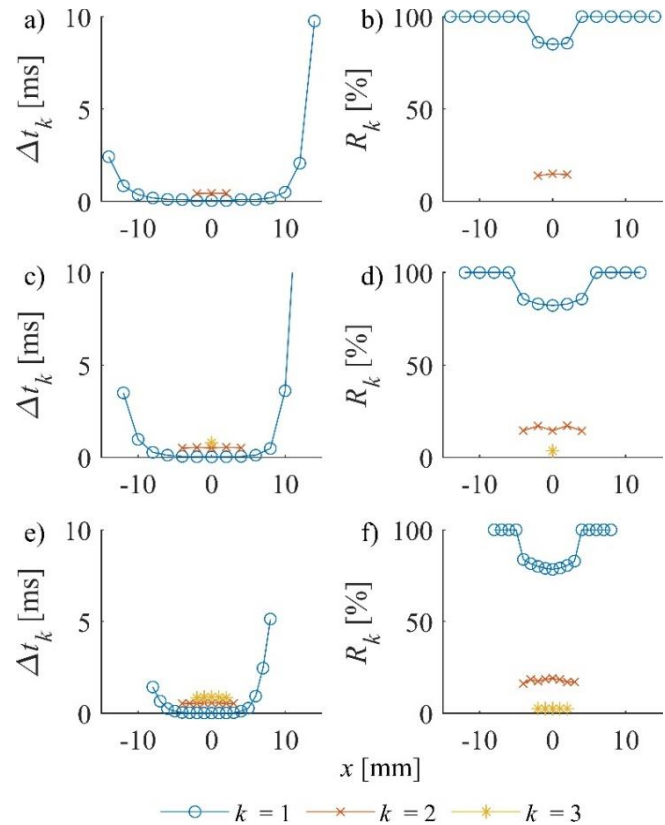

**Fig. 5.** The radial distribution (along X axis) of $\Delta t_k$ in the left column and $R_k$ in the right column at different nozzle distances: a–b) at $z = 60$ mm, c–d) at $z = 40$ mm, and e–f) at $z = 20$ mm. The results correspond to the measurement of diesel at $T_p$ = 25 °C and $p_g$ = 0.3 bar. Note that the missing marker indicates the absence of the corresponding cluster.

As the three clusters show different behavior, they are discussed separately. Figure 6 shows the radial distribution of $\Delta t_1$ and $R_1$ results of the first clusters by $p_g$ for diesel at $T_p$ = 25 °C along the X-axis. Figures 6a, 6c, and 6e show that $\Delta t_1$ shows no dependency on $p_g$ at all $z$. Due to the increasing velocity towards the central axis where enhanced atomization occurs [17, 18], there is a dependency on $r$. Figures 6b, 6d, and 6f show that $R_1$ decreases with increasing $p_g$, where more clusters appear. This effect strengthens with decreasing $z$ and $|r|$. In addition, $T_p$ does not affect either $\Delta t_1$ or $R_1$. These conclusions align with the fact that with increasing $p_g$, $V$ increases.

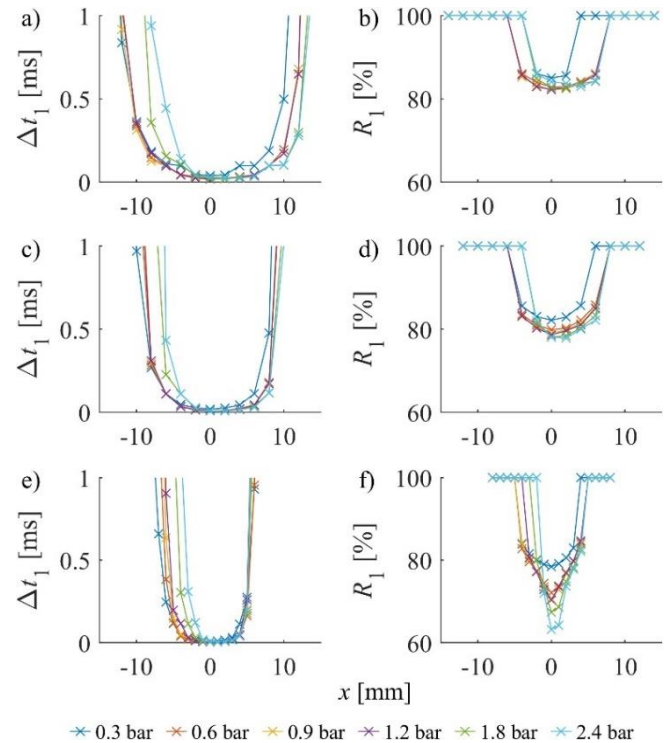

**Fig. 6.** Radial distribution (along X axis) of the results of the first cluster by $p_g$ for diesel at $T_p$ = 25 °C: first column for $\Delta t_1$ and the second column for $R_1$, while a–b) at $z$ = 60 mm, at $z$ = 40 mm, and e–f) at $z$ = 20 mm. Note that the vertical axes on a), c), and e) are limited to 1 ms to ensure transparency.

Figure 7 shows the radial distribution of $\Delta t_2$ and $R_2$ of the second clusters by $p_g$ for diesel at $T_p$ = 25 °C along the X axis. $R_2$ increases with increasing pressure, while $\Delta t_2$ shows no dependency on $p_g$. Both $\Delta t_2$ and $R_2$ decrease with increasing $z$ distances. In addition, $T_p$ does not affect either $\Delta t_2$ or $R_2$.

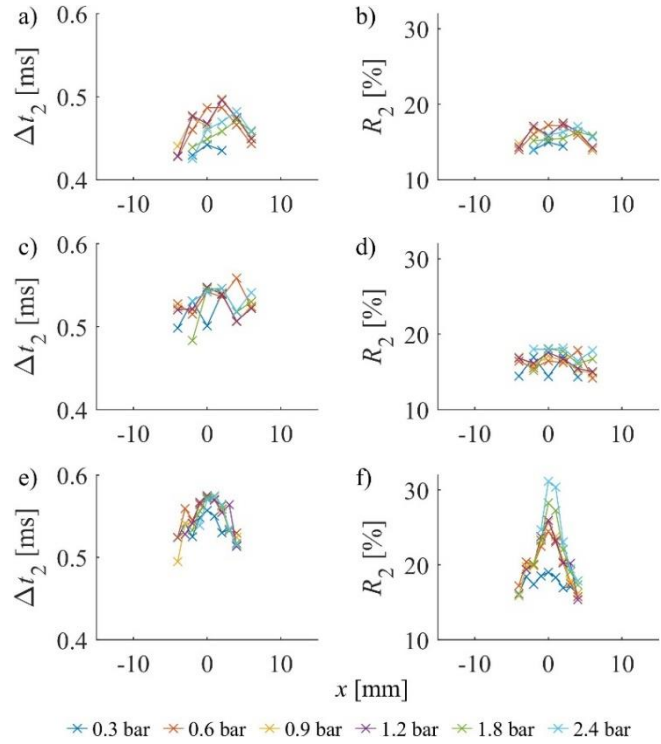

**Fig. 7.** Radial distribution (along X axis) of the results of the second cluster by $p_g$ for diesel at $T_p$ = 25 °C: first column for $\Delta t_2$ and the second column for $R_2$, while a–b) at $z$ = 60 mm, at $z$ = 40 mm, and e–f) at $z$ = 20 mm.

Figure 8 shows the radial distribution of $\Delta t_3$ and $R_3$ of the third clusters by $p_g$ for diesel at $T_p$ = 25 °C along the X-axis. Figures 8e and 8f show that $\Delta t_3$ and $R_3$ increase only at $z$ = 20 mm with increasing pressure due to enhanced atomization in the spray core. Both $\Delta t_3$ and $R_3$ have lower values with increasing $z$ distances. In addition, $T_p$ does not affect either $\Delta t_3$ or $R_3$. In conclusion, while $R_1$ decreases, $R_2$ and $R_3$ increase. Also, the radial ranges where more clusters appear may vary depending on the atomization parameters. Typically, this range at $p_g$ = 0.3 bar is narrower, as shown in Figures 6–8. These conclusions also match the fact that with increasing $p_g$, $V$ increases due to the increased clustering tendency.

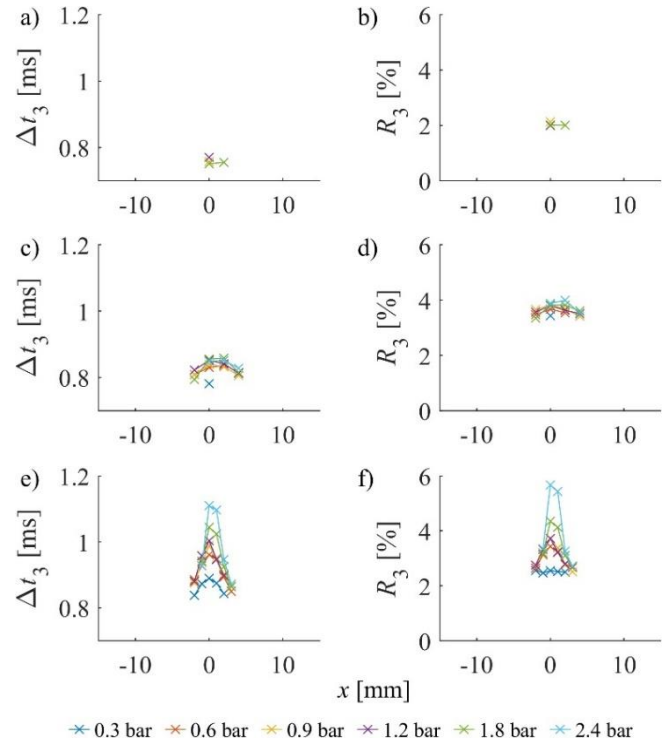

**Fig. 8.** Radial distribution (along X axis) of the results of the third cluster by $p_g$ for diesel at $T_p$ = 25 °C: first column for $\Delta t_3$ and the second column for $R_3$, while a–b) at $z$ = 60 mm, at $z$ = 40 mm, and e–f) at $z$ = 20 mm.

*3.4 Spray steadiness investigation by additional cluster removal*

To determine whether unsteadiness is natural or a result of clustering, the statistical investigation conducted in Section 3.2 was performed after removing the additional clusters ($k$ = 1 and 2) from the datasets. Figure 9 shows $V$ as a function of $p_g$ and $T_p$ along the X axis for diesel at different $z$. The markers indicate measurement points, which have more than one cluster. In this area, $V$ decreased from 0.3 to 0.05, which means a weak effect size, and unsteadiness is practically negligible. At every measurement parameter, the same trend was discovered.

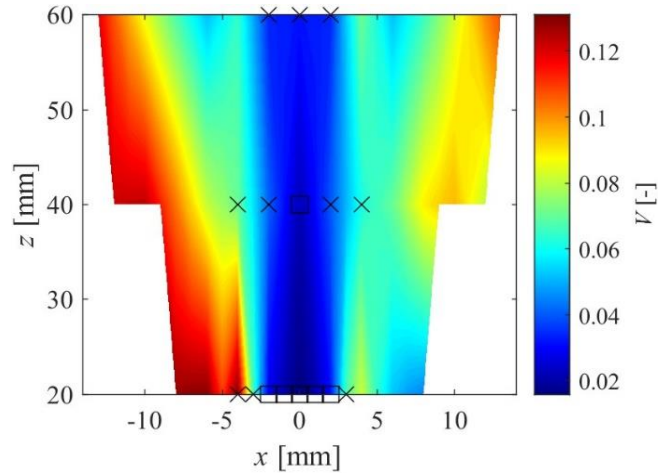

**Fig. 9.** Results of $V$ on the $x$–$z$ plane at $p_g = 0.3$ bar for diesel at $T_p = 25$ °C after removing the additional clusters. The measurement points marked with "x" originally have $k = 2$, while the markers "□" have $k = 3$.

Figure 10 shows $V$ depending on $T_p$ and $p_g$ along the X-axis for diesel after removal of additional clusters. Markers indicate measurement points with $k = 2$ or 3. In these points, $V$ decreased significantly at every $|r|$. With increasing pressure, the $V$ decreases, while there is no dependency on $T_p$. In conclusion, near the axis at $x$ and $y = 0$ mm, unsteadiness is caused by clustering, whereas at higher $|r|$, it is due to enhanced mixing and droplet entrainment [1,39].

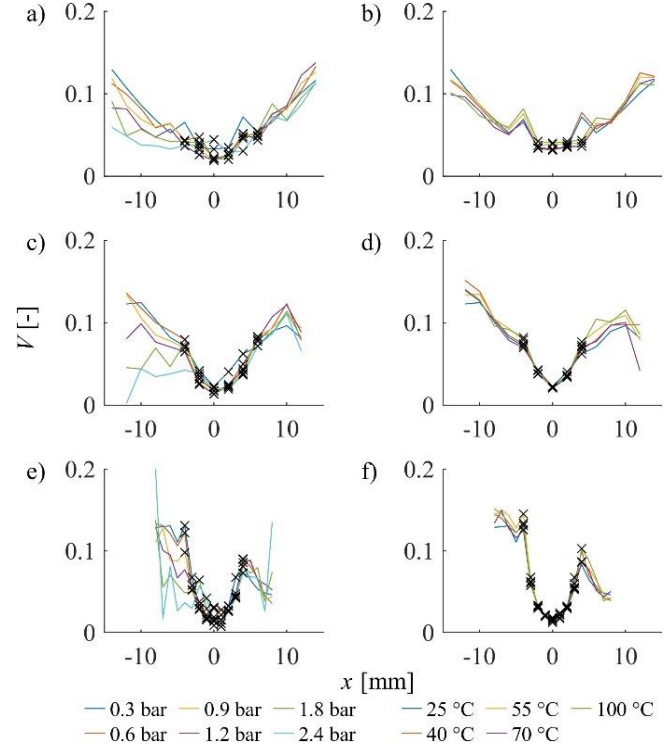

**Fig. 10.** $V$ depending on $T_p$ and $p_g$ along X axis for diesel after removal of additional clusters: first column depending on $p_g$ at $T_p = 25$ °C and second column depending on $T_p$ at $p_g = 0.3$ bar, a–b) at $z = 60$ mm, c–d) at $z = 40$ mm, and e–f) at $z = 20$ mm. Marker x indicates the presence of clustering.

*3.5 Comparison of liquids*

Table 2 presents the number of one-, two-, and three-cluster cases out of the 2,700 measurement datasets for each liquid. The one-cluster data is the most frequent case, followed by the two-cluster case and then the three-cluster case for diesel, RO, and water. However, for LHO, the three-cluster case is more frequent than the two-cluster case. Additionally, the liquids are clustered from most to least frequently as LHO, diesel, RO, and water. Hence, we can conclude that the two-cluster case transitions between the one- and three-cluster cases along the radial direction. 30% of the 10,800 datasets exhibited a clustering tendency.

**Table 2.** Number of one-, two-, and three-cluster cases by liquids.

| Liquid | $k = 1$ | $k = 2$ | $k = 3$ |
| --- | --- | --- | --- |

|       | [-]  | [%]   | [-] | [%]   | [-] | [%]   |
|-------|------|-------|-----|-------|-----|-------|
| diesel| 1583 | 58.63 | 607 | 22.48 | 508 | 18.81 |
| LHO   | 1521 | 56.33 | 582 | 21.56 | 597 | 22.11 |
| RO    | 1600 | 59.26 | 565 | 20.93 | 535 | 19.81 |
| water | 1761 | 65.22 | 523 | 19.37 | 416 | 15.41 |

In the various cases, the $R_k$ and $\Delta t_k$ statistics may differ; therefore, the results are discussed separately for each case. In the one-cluster case, $R_1$ is 100%. In the spray center, $\Delta t_1$ is close to zero with a minimum value of 0.0001 ms set by the measurement system. In contrast, towards the spray periphery, it increases, as shown in Fig. 5. $\Delta t_1$ varies significantly due to the high values at the spray periphery resulting from the low data rate and weak droplet entrainment.

Figure 11 shows the histograms of $\Delta t_k$, and Fig. 12 shows the histograms of $R_k$ grouped by liquids for the two-cluster cases. The results for the four liquids are similar, despite a difference in their clustering tendencies, as shown in Table 2. The mean value of $\Delta t_1$ is 0.034–0.036 ms, and the mean value of $\Delta t_2$ is 0.492–0.497 ms except for water, which has a 0.477 ms mean value. The mean value of $R_1$ is 82.97–84.73%, while the mean value of $R_2$ is 15.27–17.03%. The lowest $R_1$ and highest $R_2$ belong to RO, while the highest $R_1$ and lowest $R_2$ belong to water.

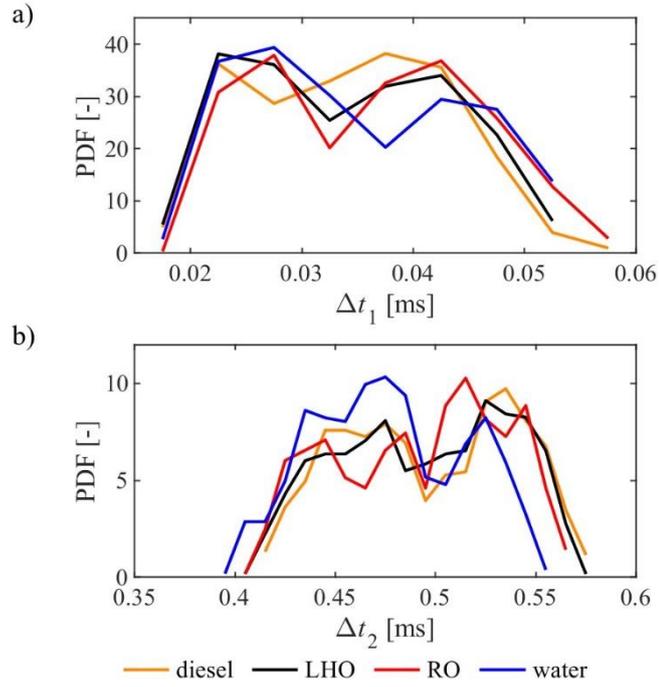

**Fig. 11.** Histogram of $\Delta t_k$ of the two-cluster cases by liquids: a) $\Delta t_1$ and b) $\Delta t_2$.

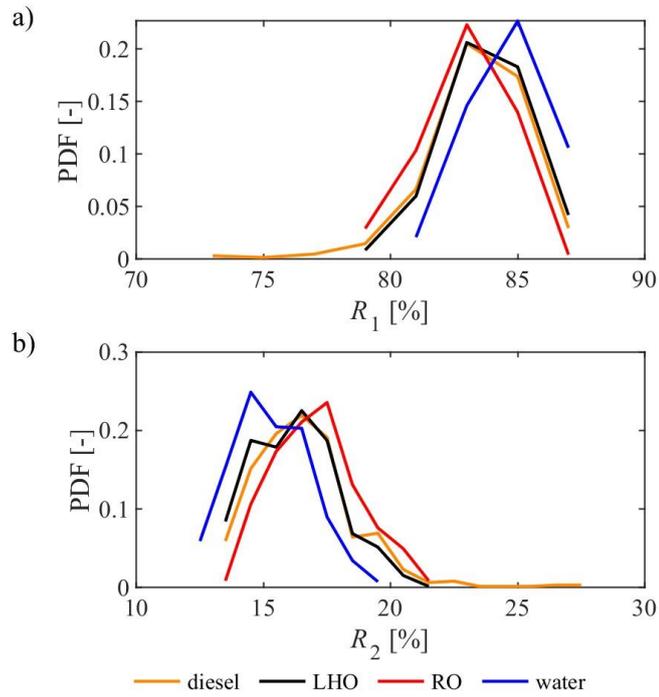

**Fig. 12.** Histogram of $R_k$ of the two-cluster cases by liquids: a) $R_1$ and b) $R_2$.

Figure 13 shows the histograms of $\Delta t_k$, and Fig. 14 shows the histograms of $R_k$ grouped by liquids for the three-cluster case. The results for the four liquids are similar to those in the two-cluster case. The mean value of $\Delta t_1$ is 0.010–0.012 ms, the mean value of $\Delta t_2$ is 0.535–0.543 ms, and the mean value of $\Delta t_3$ is 0.86–0.90 ms. $\Delta t_1$ is lower in the three-cluster case than in the two-cluster case because $k$ increases with $p_g$, and increasing velocity results in lower interparticle arrival times of droplets. On the other hand, the mean of $\Delta t_2$ is higher in the three-cluster case than in the two-cluster case. This happens because two-cluster cases appear at a higher $|r|$ region compared to the three-cluster cases, and in this region, $\Delta t_2$ decreases. The mean value of $R_1$ is 76.24–80.17%, the mean value of $R_2$ is 17.23–20.55%, and the mean value of $R_3$ is 2.6–3.21%. As three-cluster cases usually appear at higher $p_g$, $R_2$ is higher in this case compared to the two-cluster case. Because of this effect, $R_1$ is lower compared to the two-cluster case. Further related statistics of $\Delta t_k$ and $R_k$, such as minimum, maximum, mean, and standard deviation, can be found in the Supplementary Material, in Tables S1 and S2 for the two-cluster case, and in Tables S3 and S4 for the three-cluster case.

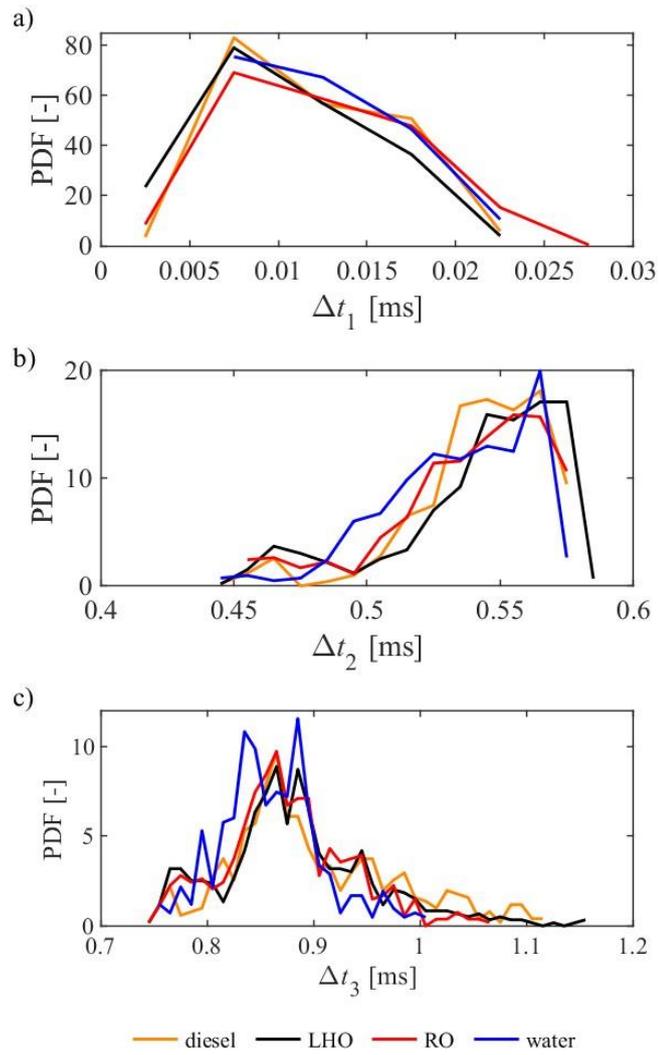

**Fig. 13.** Histogram of $\Delta t_k$ of the two-cluster cases by liquids: a) $\Delta t_1$, b) $\Delta t_2$, and c) $\Delta t_3$.

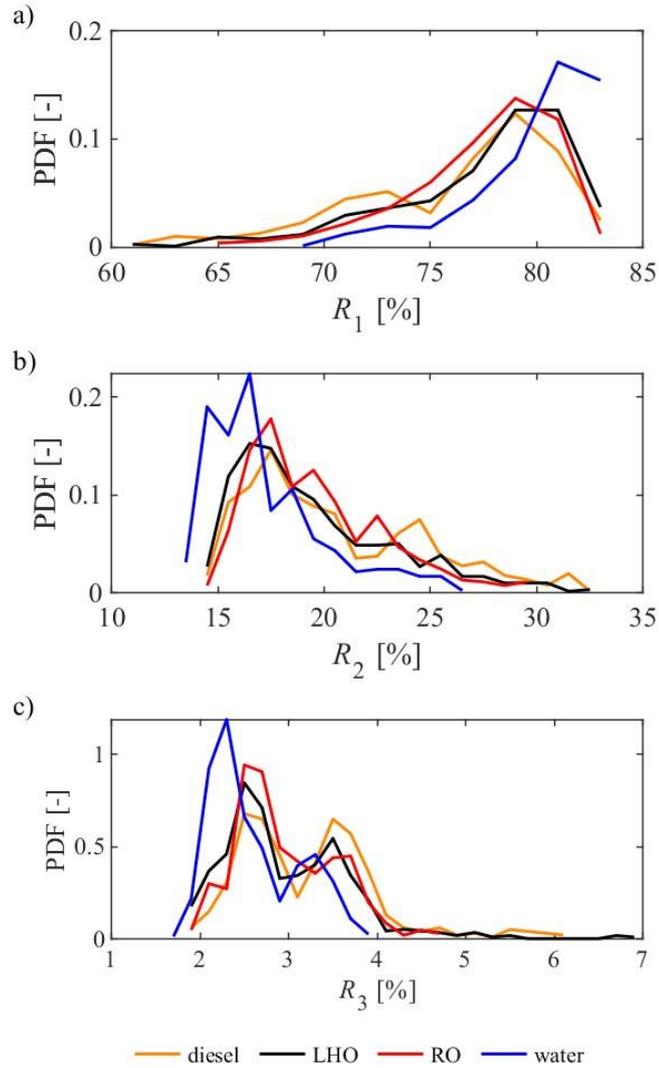

**Fig. 14.** Histogram of $R_k$ of the two-cluster cases by liquids: a) $R_1$, b) $R_2$, and c) $R_3$.

For water, clustering is less likely to occur, and as Figs. 9–12 shows the histogram is slightly different from the behavior of the other liquids, since higher peaks characterize the $R_k$ distributions. Specifically, higher $R_1$ and lower $R_2$ and $R_3$ values occur, which implies that clustering is slightly weaker for water than for other liquids. An explanation for this is the surface tension, a governing parameter of the atomization process, which is nearly three times higher for water than for other liquids [34]. This is reflected in the fact that cluster formation for water is less frequent at low pressures in the central region, while the results are similar to those of other liquids at high pressures.

Overall, the cluster formation concerns 30% of the datasets for all liquids. In these datasets, as indicated by the mean and maximum of $R_k$, a significant number of droplets belong to the second and third clusters; therefore, the clustering effect should be considered in a realistic atomizer model.

*3.6 Cluster Characterization*

The dynamic behavior of droplet clusters can be compared by investigating their $D$ and $v$ distributions. As Fig. 15 shows, the correlations of $D$, $v$, and $\Delta t$ by clusters depend on the radial position in the spray. Figures 15b and 15c show that the clusters are clearly separable. Additionally, Fig. 15a illustrates that the clusters in the $D$–$v$ plane are not separable and may differ from each other and from the characteristics of the complete dataset.

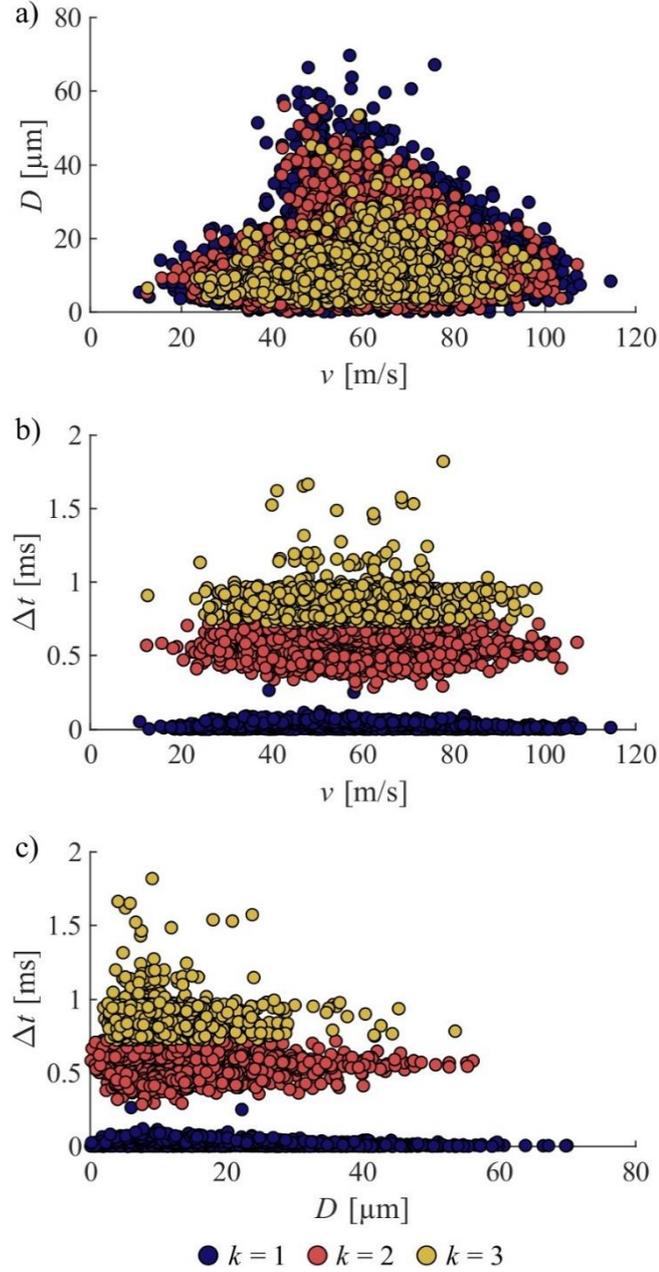

**Fig. 15.** Droplet size, velocity, and interparticle arrival time correlation for diesel at $T_p = 25$ °C, $p_g = 0.3$ bar, $z = 20$ mm, and $x = 0$ mm: a) $v$–$D$ correlation, b) $v$–$\Delta t$ correlation, and c) $D$–$\Delta t$ correlation.

Figure 16 shows the box plots of the three clusters for diesel at $T_p = 25$ °C, $p_g = 0.3$ bar, $z = 20$ mm, and $x = 0$ mm. The results show that the statistics of the clusters are similar for both $D$ and $v$ distributions. Additionally, the probability density functions plotted alongside the boxplots are similar; the differences in the minima and maxima are due to the varying $R_k$ of clusters and the inherent variance of the data. This behavior was observed in all two- and three-cluster datasets.

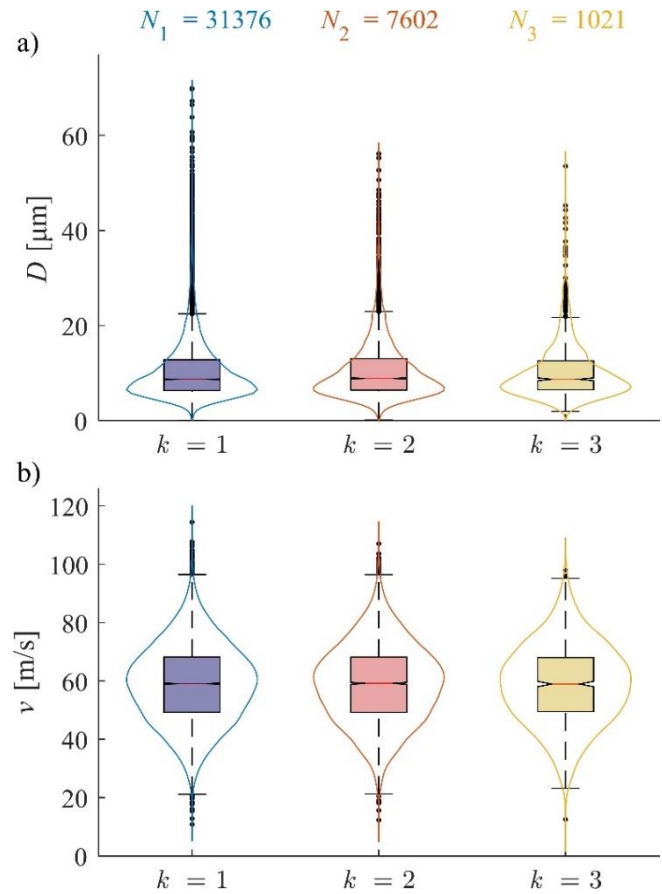

**Fig. 14.** Box plots and probability density functions of a) velocity and b) size distributions by clusters for diesel at $T_p$ = 25 °C, $p_g$ = 0.3 bar, $z$ = 20 mm, and $x$ = 0 mm.

Figure 16 shows the radial distribution of the mean value and the standard deviation of $D$ and $v$ by clusters. The results show that the statistics of $D$ and $v$ of different clusters are the same in radial positions. These conclusions correspond to all investigated measurement parameters and positions. The radial characteristics are consistent for both X and Y axes. Consequently, the velocity and droplet size distributions of the droplet clusters match the distribution of the complete realization in each spray position. Therefore, the dynamical behavior of the clusters can be modeled by the characteristics of the complete dataset, considering the corresponding $R_k$ of the clusters and $(r, z)$ position in the spray.

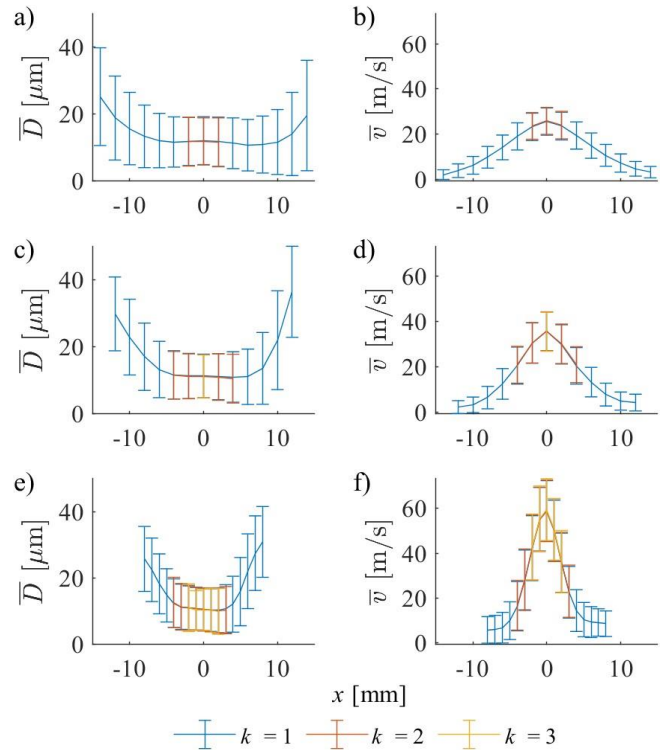

**Fig. 16.** Radial distributions of mean droplet size in the left column and mean velocity in the right column at different downstream distances: a–b) $z = 60$ mm, c–d) at $z = 40$ mm, and e–f) at $z = 20$ mm. The results correspond to the measurement of diesel at $T_p = 25$ °C and $p_g = 0.3$ bar. The error bars represent the standard deviation of the droplet size and velocity distributions.

## 4. Conclusions

The introduced clustering method was implemented for the PDA data of an airblast atomizer under a wide range of measurement conditions: four liquids at five preheating temperatures and six atomizing pressures, as well as in ninety different spray positions, totalling 10,800 data sets. The presented method can be applied generally and independently of the atomization parameters and the position of the spray:

- Clustering appears in the vicinity of the $r = 0$ mm axis, which significantly distorts the $\Delta t$ distributions from the theoretical function. The measurement point in these regions showed a high $V$ value during the $\chi^2$ test.

- With the $k$-means clustering, the determined clusters can be removed from the experimental datasets. In these filtered cases, the $V$ value decreased below ~0.05, which indicates practically negligible unsteady behavior. In conclusion, at these measurement points, the unsteadiness is caused by the clustering effect. In contrast, in the peripheral region of the spray, the unsteadiness is caused by mixing and droplet entrainment.

- Cluster formation concerns a significant fraction of the data sets (~30%). Up to 40% of the droplets of these datasets show clustering. Therefore, the effect of this process is significant for modeling.

- The centroids of the clusters have typical interparticle arrival time values, shown in Tables S1 and S3. These values are independent of $T_p$, but they depend on $p_g$, $z$, and $r$.

- The number of clusters, $k$, depends on the radial and axial position and the atomizing pressure and does not correlate with the preheating temperature.

- The droplet ratio of the first cluster decreases, while that of the second and third clusters increases with increasing atomizing pressure.

- For diesel, LHO, and RO, the results are similar, while for water, clustering is less frequent, and lower $\Delta t_k$ characterizes the clusters, as the surface tension is nearly three times higher for water than for other liquids.

- The velocity and droplet size distributions of the droplet clusters match the distribution of the complete realization at all investigated parameters. Consequently, the dynamical behavior of the clusters can be modeled by the characteristics of the complete dataset, considering the corresponding droplet ratio present in the clusters.

Assuming that additional clusters are superimposed over a quasi-steady process, the irregularities can be incorporated into a time-series spray model by modeling the behavior of the clusters individually. Therefore, further investigation of the randomness and spatial distribution of the determined clusters is the subject of a forthcoming work.

**Supplementary Material**

Further related statistics of $\Delta t_k$ and $R_k$, such as minimum, maximum, mean, and standard deviation, can be found in the Supplementary Material, in Tables S1 and S2 for the two-cluster case, and in Tables S3 and S4 for the three-cluster case.

**Acknowledgements**

This project was supported by the National Research. Development and Innovation Fund of Hungary, project No.s OTKA-FK 137758 and NKKP ADVANCED 150696, the János Bolyai scholarship of the Hungarian Academy of Sciences, the University Research Fellowship Programme (EKÖP-24–3-BME-155, EKÖP-25-3-BME-246) and the Doctoral Excellence Fellowship Programme (DKÖP-23–1-BME-55) funded by the National Research Development and Innovation Fund of the Ministry of Culture and Innovation of Hungary and the Budapest University of Technology and Economics, under a grant agreement with the National Research, Development and Innovation Office. The authors acknowledge the financial support from the Czech Science Foundation grant No. 25-17759S and from the project "Mechanical Engineering of Biological and Bio-inspired Systems", funded as project No. CZ.02.01.01/00/22_008/0004634 by Programme Johannes Amos Commenius.

**Conflict of Interest Statement**

The authors have no conflicts to disclose.

**Data Availability Statement**

The data that support the findings of this study are openly available in Zenodo at http://doi.org/10.5281/zenodo.17935932.

**Author Contributions**

**Erika Rácz**: Conceptualization, Methodology, Software, Validation, Formal analysis, Data curation, Writing - Original Draft, Visualization.

**Milan Malý**: Validation, Investigation, Resources, Data curation, Supervision.

**Jan Jedelský**: Resources, Supervision, Project administration, Funding acquisition.

**Viktor Józsa**: Conceptualization, Validation, Writing - Original Draft, Supervision, Project administration, Funding acquisition.